\documentclass[showpacs,floatfix,amsmath,amssymb,nofootinbib]{revtex4}
\usepackage[T1]{fontenc}
\usepackage{slashed}
\usepackage{bm}
\usepackage{dcolumn}
\usepackage{graphicx}
\usepackage{bbold}

\newcommand{\beq}{\begin{equation}}
\newcommand{\eeq}{\end{equation}}
\newcommand{\beqa}{\begin{eqnarray}}
\newcommand{\eeqa}{\end{eqnarray}}
\begin{document}

\title
{\bf Polarization observables for elastic electron scattering off a
  moving nucleon }
\author{H. Arenh\"ovel}
\affiliation{Institut f\"ur Kernphysik, Johannes Gutenberg-Universit\"at Mainz, D-55099
Mainz, Germany}
\date{\today}

\begin{abstract}
General expressions for all parity-conserving polarization observables
of elastic electron-nucleon scattering 
in the one-photon exchange approximation are derived for a general frame of
reference, i.e.\ not assumming  scattering off a nucleon at
rest and not specializing to a specific system of
coordinates. Essentially, the given expressions are also valid for the
inverse process, i.e.\ nucleon scattering off electrons. 
\end{abstract}

\pacs{ 25.30.Bf, 13.60.Fz, 13.88.+e} \maketitle

\section{Introduction}
The present study was initiated by recent experiments on quasi-elastic
electron-nucleus scattering $\vec e\,(A,A')\vec p$, measuring the
polarization transfer from an incoming longitudinally polarized
electron to 
an emitted proton~\cite{Ya17}. The measured outgoing proton
polarization was then compared to the elementary process $\vec e +p
\to e'+\vec p$ for the same squared four momentum transfer assuming
the scattering off a proton at rest. However, the bound proton,
on which the electron scatters, is not at rest at all because of the
Fermi motion of the struck proton inside the nucleus.
Thus it appeared, that a more appropriate comparison should be done 
with the elementary process on a moving proton taking as momentum for
the struck proton the negative missing momentum as an
approximation~\cite{Pa19}. Indeed, it was found that the correction
to the ratios of the transverse and longitudinal polarization
components of the emitted proton to the ones of the elementary scattering
due to the initial motion was up to 20~\% at high missing   momentum
although the correction in the double ratio with respect to the
transverse over the longitudinal components was only a few percent. 

The study of polarization observables in elastic electron-proton
scattering has a long history, both theoretical and experimental. Early
theoretical studies are found in Refs.~\cite{Bi57,Ra57,Fo57,Sc59,Ak68} and
early reviews in~\cite{Do69,Ak73}, where specific types of
polarization observable were considered. Usual special reference
frames have been chosen for the consideration of specific polarization
observables. More recently, Gakh et al.~\cite{Ga11}
have studied polarization effects in elastic proton-electron
scattering, which on a formal basis is equivalent to electron-proton
scattering. They have considered three types of polarization 
observables: outgoing proton polarization transferred either from a
polarized target electron or from an incoming polarized proton, and
the beam target asymmetry from polarized beam and target. As
reference frame is taken the one, in which the target electron is at rest, and
they use a special coordinate system related to the scattering
kinematics with respect to the notation of Ref.~\cite{By78}. 

It is the aim of the present work to derive explicit expressions for
all possible parity conserving observables 
in a general form without choosing a special frame of reference nor a
special coordinate system so
that one can easily evaluate any observable for an arbitrary reference
frame and an arbitrary coordinate system. For that purpose we also
introduce an intuitive and compact general notation.

In the next Section~\ref{definition} we introduce a general definition
of an observable of the scattering process as a trace over the spin
degrees of freedom of the initial and final  states of a hermitean
quadratic form in the $T$-matrix elements and associated density
matrices. Within the one-photon exchange approximation any observables
is then given as a contraction of a corresponding lepton tensor with
a hadron tensor. Their specific forms depend on the type observables
and are presented in Section~\ref{tensors}. Section~\ref{obs} is
devoted to explicit expressions of the various observables. A short
summary and conclusions are given in Section~\ref{conclusion}. 
Some details are contained in three appendices. In Appendix~\ref{app1}
the derivation of the lepton and hadron tensors are sketched, and
Appendix~\ref{app2} lists explicit expressions of the beam-target
asymmetries of the final spin correlations. The specialization to
longitudinal polarized initial electrons for the beam-target asymmetry
and the electron-nucleon polarization transfer with explicit
expressions for the lab and Breit frames is presented in
Appendix~\ref{app3}.

\section{General definition of an observable\label{definition}}
In elastic electron-nucleon scattering
\beq
e(k) + N(p) \to e'(k') + N'(p')\,,
\eeq
where $k=(k_0,\vec k\,)$ and $k'=(k'_0,\vec k^{\,\prime})$ stand for
the four-momenta of incoming and scattered electron, respectively, and $p=
(E_p,\vec p\,)$ and $p'=(E_{p'},\vec p\,^{\,\prime})$ for the
corresponding quantities of the nucleon, any observable $\cal O$, for
example, unpolarized scattering cross section, beam and target asymmetries,
polarization transfer from polarized incoming electron to 
the final nucleon, etc., is defined by~\cite{Ak73}
\beq
{\cal O} \frac{d\sigma_0}{d\Omega_{e'}}=F_{kin}
\, \Sigma_{fi}({\cal O})\,,\label{observable}
\eeq
where $d\sigma_0/d\Omega_{e'}$ denotes the unpolarized
differential cross section. The quantity $\Sigma_{fi}({\cal O})$ depends on the type 
of observable $\cal O$ and is given as a trace over all
spin degrees of freedom of initial and final electron and nucleon, as
indicated by the superscript $S$ at the trace symbol $Tr^S$ 
\beq
\Sigma_{fi}({\cal O})=\frac{Q^4}{4\alpha^2}Tr^S(T^{fi} \rho_i({\cal O})T^{fi,\dagger}
\rho_f({\cal O}))\label{trace}\,.
\eeq
Here, $T^{fi}$ denotes the $T$-matrix of the scattering process
and $\rho^{i/f}({\cal O})$ denote the spin density matrices of the
initial and final states, which
depend on the observable ${\cal O}$, whether it involves polarized
or unpolarized initial particles and whether the polarization of
the final particles is analyzed. Furthermore, $\alpha$ denotes the
fine structure constant and $Q^2=-q^2$ the squared four momentum
transfer with $q=k-k'=p'-p$. The factor $Q^4/\alpha^2$ 
has been included in view of the
one-photon exchange approximation for the $T$-matrix.

The kinematic factor has the form\footnote{The symbol $\bar v$ means 
  $|\vec v|$ where it is needed 
  to distinguish it from the four vector $v=(v_0,\vec v\,)$.}
\beqa
F_{kin}&=&\frac{2^2\alpha^2m_e^2 m_N^2 \bar k^{\prime}}{Q^4
\, \sqrt{(k\cdot p)^2-m_e^2 m_N^2}}\,\frac{1}{E_{p'}|1+dE_{p'}/dk_0'|}\nonumber\\
&=&
\frac{2^2\alpha^2m_e^2 m_N^2 \bar k^{\prime 2}}{Q^4[k\cdot p+g(p,k,k')]
\, \sqrt{(k\cdot p)^2-m_e^2 m_N^2}}\,. \label{fkin}
\eeqa
where $m_e$ and $m_N$ denote the masses of electron and nucleon,
respectively, and 
\beq
g(p,k,k')=(\bar k' -k_0')(E_{p'}-\bar k'+\vec e_{k'}\cdot (\vec
p+\vec k))\,,
\eeq
with $\vec e_{k'}=\vec k^{\,\prime}/{\bar k'}$ denoting the unit
vector along $\vec k^{\, \prime}$.
It is worth noting that the same formal expressions apply for the inverse process, i.e.\
nucleon-electron scattering~\cite{Ga11}. One only has to exchange
$k\leftrightarrow p$, $k'\leftrightarrow p'$, 
$\Omega_{e'}\leftrightarrow \Omega_{p'}$, and $m_e\leftrightarrow m_N$.
 In the high energy limit ($m_e/k_0\approx 0$) $g(p,k,k')$ tends to
zero, and the kinematic factor becomes 
\beq
\widetilde F_{kin}=\frac{2^2\alpha^2 m_e^2 m_N^2 \bar k^{\prime 2}}{
Q^4\,k\cdot p\,\sqrt{(k\cdot p)^2-m_e^2 m_N^2}}\,.\label{fkin_he}
\eeq

In the one-photon exchange approximation, used throughout in this work,
the $T$-matrix is given by 
the contraction of the leptonic current $J_e$ and the hadronic one
$J_N$, i.e.\footnote{We use the Einstein convention for summations over
  greek indices of four-vectors and four-tensors.}
\beq
T^{fi}_{s_{e'}, s_{N'}, s_{e}, s_{N}}=\frac{\alpha}{Q^2}\,
J_{e,\mu}(k',s_{e'};k,s_{e})
J_N^\mu (p',s_{N'};p,s_{N})\,.\label{tmatrix}
\eeq

The spin density matrices $\rho^{i/f}({\cal O})$ are given as products
of electron and nucleon spin operators depending on the rest
frame spins $\vec s_{e}^{\, i/f}$ and $\vec s_{N}^{\, i/f}$ 
of initial and final electrons and nucleons, respectively,
i.e.\
\beqa
\rho^{i}(k,\vec s_e^{\, i};p, \vec s_N^{\, i})({\cal O})&=&
\rho^{i}_e(k,\vec s_e^{\, i};{\cal O})\,\rho^{i}_N(p,\vec s_N^{\, i};{\cal O})
\,,\label{rho_i}\\
\rho^{f}(k',\vec s_{e}^{\, f};p', \vec s_{N}^{\, f})({\cal O})&=&
\rho^{f}_e(k',\vec s_{e}^{\, f};{\cal O}) \,\rho^{f}_N(p',\vec s_{N}^{\, f};{\cal O})
\,. \label{rho_f}
\eeqa
Their specific form depends on the polarization state of the
corresponding particle, i.e.\ unpolarized or polarized as required by
the specific observable. For example, for the density
matrix of the initial electron  
\beq
\rho^i_e (k,\vec s_e ^{\, i}; {\cal O})_{s'_e s_{e}} =
\bar u_e(k, s'_e) {\cal S}^i_e ({\cal O}) u_e(k,s_e)\label{rho_ei}
\eeq
with $u_e(k,s)$ as electron Dirac spinor, one has two possibilities
for the Dirac operator ${\cal S}^i_e ( {\cal O})$, namely
\beq
{\cal S}^i_e ( {\cal O}) =
\left\{
\begin{array}{ll}
\mathbb{1}_4 & \mathrm{unpolarized}\\
\gamma_5\slashed{S}^{\, i}_e (k,\vec s_e ^{\, i}\,) &
\mathrm{polarized}\\
\end{array}\right\}\,.\label{rho_op_e}
\eeq
The relativistic spin four vector $S ^{\, i}_e (k,\vec s_e ^{\, i}\,)$ is
related to the spin three vector $\vec s^{\, i}$ in the electron's rest frame by  
\beq
S_e ^{\, i} (k,\vec s_e ^{\, i}\,)=\Big(\frac{\vec s_e ^{\, i}
\cdot \vec k }{m_e}, \vec s_e ^{\, i}+\frac{\vec s_e ^{\, i}\cdot
  \vec k }{m_e(k_0+m_e)}\vec k\,\Big)\,.
\label{spin}
\eeq
It is obtained from the electron's spin four vector 
$S_e^{\, i} (0,\vec s_e ^{\, i}\,)=(0,\vec s_e ^{\, i}\,)$ in
the rest system by a
Lorentz boost $L(\vec\beta\,)$, i.e.
\beq
S_e ^{\, i} (k,\vec s_e ^{\, i}\,)=L(\vec\beta\,)\, S_e ^{\, i} (0,\vec s_e ^{\, i})
\eeq
where $\vec\beta=\vec k/k_0$.
The relativistic spin operator obeys the following properties
\beqa
S_e(k, \vec s_e\,)\cdot S_e(k,\vec s_e\,)&=&-1\,,\label{espina}\\
S_e(k,\vec s_e\,)\cdot k&=&0\,. \label{espinb}
\eeqa
 Corresponding expressions hold for the densitiy matrices of the
final electron with the spin operator $S_e^f$ and for the initial and final nucleons
with  $S_N^{i/f}$, respectively, as function of the observable ${\cal O}$. 

In view of the separation of the $T$-matrix (Eq.~(\ref{tmatrix})) and
the density matrices (Eqs.~(\ref{rho_i}) and (\ref{rho_f})) into a
leptonic and a hadronic part, the trace $\Sigma_{fi}({\cal O})$  
can be represented as 
the contraction of a leptonic tensor $\eta_{\mu\nu}^{\cal O} (J_e^{fi})$ and a
hadronic one $\eta_{\mu\nu}^{\cal O} (J_N^{fi})$
\beq
\Sigma_{fi}({\cal O})=\eta_{\mu\nu}^{\cal O} (J_e^{fi}) 
\eta^{{\cal O},\mu\nu}(J_N^{fi})\,,
\eeq
where the tensors are given as traces over the corresponding spin
degrees of freedom, i.e.\ for $a\in\{e,N\}$
\beq
\eta_{\mu\nu}^{\cal O} (J_a^{fi})=\frac{1}{2}\,
Tr^S\Big(J^{fi}_{a,\mu}\rho^i_{a}({\cal O})
J^{fi,\dagger}_{a,\nu}\rho^f_{a}({\cal O})\Big)\,.\label{tensor}
\eeq

\section{Lepton and hadron tensors\label{tensors}}

According to the two possibilities for the density matrices of initial
and final particles, i.e.\ whether they are unpolarized or polarized
(see Eq.~(\ref{rho_op_e})), one finds four  
types of tensors for both the lepton and the hadron sector, namely
initial and final particles unpolarized, one initial or final particle
polarized and both particles polarized. The derivation of these
tensors is sketched in Appendix~\ref{app1}. 

Here we list the resulting explicit expressions, where 
we introduce for convenience as a shorthand for the relativistic
spin four-vectors of initial and final electrons
\beq
S_{e}^{i/f}=S_e(k/k',\vec s_e^{\, i/f})\,,\label{notation-sif}
\eeq
and corresponding notations for the nucleon spin vectors $(S_{N}^{i/f})$. If the spin
three-vector in the particle's rest frame $\vec s^{\,i/f}$ has to be specified we use 
$S_{e}^{i/f}(\vec s ^{\,i/f}\,)$. Furthermore, we introduce two symmetric four-tensors
\beqa
\Omega_{\mu\nu}&=&q^2g_{\mu\nu}-q_\mu q_\nu\,,\label{qmunu}\\
\Sigma_{a,2;\mu\nu}&=&S^i_{a;\mu} S^f_{a;\nu}+ S^i_{a;\nu} S^f_{a;\mu}\label{smunu}\,,
\eeqa
with $a\in \{e,N\}$, and the following spin dependent scalars 
\beqa
\Sigma_{a,0}&=&\frac{1}{2}\Sigma_{a,2;\mu}^{\mu} 
=S_{a}^i \cdot S_{a}^f \,, \label{s0}\\
\Sigma_{a,2}(v)&=&\frac{1}{2} v_\mu\Sigma_{a,2}^{\mu\nu} v_\nu
=v\cdot S_{a}^i \, v\cdot S_{a}^f \,,\label{s2}
\eeqa
for a four-vector $v$.
Then, as shown in Appendix~\ref{app1}, one obtains with $P=p+p'$ and
$q=p'-p$ for the hadron tensors  
\beqa
\eta_{\mu\nu}^{N,0}(p',p)&=&\frac{1}{4m_N^2}
\Big(\frac{G_E^2+\tau G_M^2}{1+\tau}\,P_\mu P_\nu
+G_M^2\Omega_{\mu\nu}\Big) \,,\label{p-tensor-0}\\
\eta_{\mu\nu}^{N,\vec s^{\,i/f}_N}(p',p)  &=& -\frac{iG_M}{2m_N} 
\Big[\frac{G_E- G_M}{4m_N^2 (1+\tau )}\,
\Big (P_\mu\epsilon_{\nu \alpha \beta \gamma}
-(\mu\leftrightarrow\nu) \Big)P^\gamma
+G_M\epsilon_{\mu\nu \alpha\beta} 
\Big]S_{N}^{i/f;\alpha} q^\beta\,,
\label{p-tensor-s}\\
\eta_{\mu\nu}^{N,\vec s^{\, i}_N, \vec s^{\, f}_N}(p',p)  &=& 
\frac{G_M^2}{4m_N^2}\,\Big[2\Sigma_{N,2}(q)g_{\mu\nu}
-\Sigma_{N,0} (\Omega_{\mu\nu} + P_\mu P_\nu)\nonumber\\
&&+q^2\, \Sigma_{N,2;\mu\nu}
+\Big((P_\mu \Sigma_{N,2;\nu\rho}P^\rho
-q_\mu \Sigma_{N,2;\nu\rho} q^\rho) +(\mu
\leftrightarrow \nu)\Big)\Big]\nonumber\\
&&+\frac{G_M(G_E-G_M)}{4m_N^2(1+\tau)}\,\Big[
(P_\mu \Sigma_{N,2;\nu\rho} P^\rho +(\mu\leftrightarrow\nu))
-2\Sigma_{N,0}\,P_\mu P_\nu\Big]\nonumber\\
&&-\frac{(G_E-G_M)^2}{16m_N^4(1+\tau)^2}\, \Big (2\Sigma_{N,2} (q)
+\Sigma_{N,0}P^2 \Big) P_\mu P_\nu \,, \label{p-tensor-ssa}
\eeqa
where $\epsilon_{\mu\nu\alpha\beta}$ denotes the
four dimensional totally antisymmetric Levi-Civita tensor. 
The expression in Eq.~(\ref{p-tensor-ssa}) can be simplified yielding
\beqa
\eta_{\mu\nu}^{N,\vec s^{\, i}_N, \vec s^{\, f}_N}(p',p)  &=& 
\frac{1}{4m_N^2}\,\Big\{ 
G_M^2 \Big[ 2\Sigma_{N,2}(q) g_{\mu\nu} -\Sigma_{N,0} \Omega_{\mu\nu}+q^2\, \Sigma_{N,2;\mu\nu}
-\Big (q_\mu \Sigma_{N,2;\nu\rho} q^\rho 
+(\mu \leftrightarrow \nu) \Big) \Big]\nonumber\\
&&-\Big[\frac{G_E^2+\tau G_M^2}{1+\tau}\,\Sigma_{N,0}
+\frac{(G_E-G_M)^2}{2m_N^2(1+\tau)^2}\, \Sigma_{N,2}(q)\Big]
\,P_\mu P_\nu \nonumber\\
&&
+\frac{G_M(G_E+\tau G_M)}{1+\tau}\,
\Big (P_\mu \Sigma_{N,2;\nu\rho} P^\rho +(\mu\leftrightarrow\nu) \Big) \Big\}
\,. \label{p-tensor-ss}
\eeqa

The lepton tensors are obtained from the above ones by the 
replacements $p\to k$ and $p' \to k'$, i.e.\ $P\to K=k+k'$, and $q\to
-q=k'-k$, $S_{N}^{i/f}\to S_{e}^{i/f}$, and furthermore $m_N\to m_e$ and 
$G_E=G_M=1$, yielding
\beqa
\eta_{\mu\nu}^{e,0}(k',k)&=&
\frac{1}{4m_e^2}(K_\mu K_\nu +\Omega_{\mu\nu})
\label{e-tensor-0}\,,\\
\eta_{\mu\nu}^{e,\vec s^{\,i/f}_e}(k',k)&=&
\frac{i}{2m_e}\,\epsilon_{\mu\nu\alpha\beta}S_{e}^{i/f;\alpha} q^\beta\,,
\label{e-tensor-s}\\
\eta^{e,\vec s_e^{\, i}, \vec s_e^{\,f}}_{\mu\nu}(k',k)&=&
\frac{1}{4m_e^2}\,\Big\{ 
\Big[ 2\Sigma_{e,2}(q) g_{\mu\nu} -\Sigma_{e,0} (\Omega_{\mu\nu}+K_\mu K_\nu)
+q^2\, \Sigma_{e,2;\mu\nu}\nonumber\\
&&+\Big( (K_\mu \Sigma_{e,2;\nu\rho} K^\rho 
-q_\mu \Sigma_{e,2;\nu\rho} q^\rho) 
+(\mu \leftrightarrow \nu) \Big) \Big]\Big\}
\,, \label{e-tensor-ss}
\eeqa
where $S_{e}^{i/f}=S ^{i/f}_e (k/k',\vec s_e^{\,i/f})$. 

We would like to point out, that
the hadron single spin tensors $\eta_{\mu\nu}^{N,\vec s^{\,i/f}_N}(p',p)$
as well as the lepton single
spin tensors $\eta_{\mu\nu}^{e,\vec s^{\,i/f}_e}(k',k)$  formally
have the same
structure except for the replacements $S_{N}^{i}\to S_{N}^{f}$ and
$S_{e}^{i}\to S_{e}^{f}$, respectively.

\section{ Observables\label{obs}}

Now we will consider all possible observables distinguishing
between parity conserving and non-conserving ones, where the latter
ones are listed only. As mentioned above,
to each observable ${\cal O}$ is associated a pair of specific lepton and hadron 
tensors as determined by the corresponding density
matrix operators ${\cal S}^{i/f}_{e/N} ({\cal O})$ according to Eqs.~(\ref{rho_i}) through 
(\ref{rho_op_e}). Observables, density matrix operators, and tensors are listed
in Table~\ref{tab1} for the parity conserving observables and in
Table~\ref{tab2} for the parity non-conserving ones. 

\begin{table}[h]
\caption{Listing of parity conserving observables $\cal O$,
  corresponding density matrix 
  operators ${\cal S}^{i/f}_{e/N} ({\cal O})$, and tensors $\eta^{e/N}_{\mu\nu}({\cal O})$.}
\begin{ruledtabular}
\begin{tabular}{l|ll|l|ll|l}
$\cal O$ & ${\cal S}^i_e ({\cal O})$& 
 ${\cal S}^f_{e}({\cal O})$ &$\eta^e_{\mu\nu}({\cal O})$&
${\cal S}^i_N ({\cal O})$& 
${\cal S}^f_{N}({\cal O})$& $\eta^N_{\mu\nu}({\cal O})$\\
\colrule
$ 1$ & 
$\mathbb{1}_4 $ & $\mathbb{1}_4$ & $\eta^{e,0}_{\mu\nu}$&
$\mathbb{1}_4 $  &$\mathbb{1}_4 $& $\eta^{N,0}_{\mu\nu}$\\
$ A_{eN} $ &
$\gamma_5\slashed{S}_e^i$ & $\mathbb{1}_4$& $\eta^{e,\vec s_e^{\,i}}_{\mu\nu}$
& $\gamma_5\slashed{S}_N^i $ &$\mathbb{1}_4 $& $\eta^{N,\vec s_N^{\,i}}_{\mu\nu}$\\
\colrule
$ P_{e'}^e $ &
$\gamma_5\slashed{S}_e^i $ &$\gamma_5\slashed{S}_{e}^f$& $\eta^{e,\vec s_e^{\,i},\vec s_e^{\,f}}_{\mu\nu}$
& $\mathbb{1}_4 $ & $\mathbb{1}_4$& $\eta^{N,0}_{\mu\nu}$\\
$ P_{e'}^N $ &
$\mathbb{1}_4 $ &$\gamma_5\slashed{S}_{e}^f$& $\eta^{e,\vec s_e^{\,f}}_{\mu\nu}$
& $\gamma_5\slashed{S}_N^i$ & $\mathbb{1}_4$& $\eta^{N,\vec s_N^{\,i}}_{\mu\nu}$\\
\colrule
$ P_{N'}^e $ &
$\gamma_5\slashed{S}_e^i $ & $\mathbb{1}_4 $& $\eta^{e,\vec s_e^{\,i}}_{\mu\nu}$
&$\mathbb{1}_4$ &$\gamma_5\slashed{S}_{N}^f$& $\eta^ {N,\vec s_N^{\,f}}_{\mu\nu}$\\
$ P_{N'}^N $ &
$\mathbb{1}_4 $ & $\mathbb{1}_4$ & $\eta^{e,0}_{\mu\nu}$
&$\gamma_5\slashed{S}_N^i$&$\gamma_5\slashed{S}_{N}^f$& $\eta^{N,\vec s_N^{\,i}, \vec s_N^{\,f}}_{\mu\nu}$\\
\colrule
$ P_{e'N'} $ &
$\mathbb{1}_4 $ &$\gamma_5\slashed{S}_{e}^f$& $\eta^{e,\vec s_e^{\,f}}_{\mu\nu}$&
 $\mathbb{1}_4 $ &$\gamma_5\slashed{S}_{N}^f$& $\eta^{N,\vec s_N^{\,f}}_{\mu\nu}$\\
$ P_{e'N'}^{eN} $ &
$\gamma_5\slashed{S}_e^i $&$\gamma_5\slashed{S}_{e}^f$ & $\eta^{e,\vec s_e^{\,i},\vec s_e^{\,f}}_{\mu\nu}$&
 $\gamma_5\slashed{S}_N^i $&$\gamma_5\slashed{S}_{N}^f$& $\eta^{N,\vec s_N^{\,i}, \vec s_N^{\,f}}_{\mu\nu}$\\
\end{tabular}
\end{ruledtabular}
\label{tab1}
\end{table}

\begin{table}[h]
\caption{Listing of parity non-conserving observables $\cal O$,
  corresponding density matrix 
  operators ${\cal S}^{i/f}_{e/N} ({\cal O})$, and tensors $\eta^{e/N}_{\mu\nu}({\cal O})$.}
\begin{ruledtabular}
\begin{tabular}{l|ll|l|ll|l}
$\cal O$ & ${\cal S}^i_e ({\cal O})$& 
 ${\cal S}^f_{e}({\cal O})$ &$\eta^e_{\mu\nu}({\cal O})$&
${\cal S}^i_N ({\cal O})$& 
${\cal S}^f_{N}({\cal O})$& $\eta^N_{\mu\nu}({\cal O})$\\
\colrule
$ A_e $ &
$\gamma_5\slashed{S}_e^i $ & $\mathbb{1}_4$& $\eta^{e,\vec s_e^{\,i}}_{\mu\nu}$
& $\mathbb{1}_4 $ & $\mathbb{1}_4 $& $\eta^{N,0}_{\mu\nu}$\\
$ A_N $ &
$\mathbb{1}_4 $ & $\mathbb{1}_4$&  $\eta^{e,0}_{\mu\nu} $
& $\gamma_5\slashed{S}_N^i $ &$\mathbb{1}_4 $& $\eta^{N,\vec s_N^{\,i}}_{\mu\nu}$\\
\colrule
$ P_{e'} $ &
$\mathbb{1}_4 $ &  $\gamma_5\slashed{S}_{e}^f$& $\eta^{e,\vec s_e^{\,f}}_{\mu\nu}$ 
&$\mathbb{1}_4  $ &$\mathbb{1}_4$& $\eta^{N,0}_{\mu\nu}$\\
$ P_{e'}^{eN} $ &
$\gamma_5\slashed{S}_e^i $ & $\gamma_5\slashed{S}_{e}^f$&$\eta^{e,\vec s_e^{\,i},\vec s_e^{\,f}}_{\mu\nu}$
& $\gamma_5\slashed{S}_N^i $  & $\mathbb{1}_4$& $\eta^{N,\vec s_N^{\,i}}_{\mu\nu}$\\
\colrule
$ P_{N'} $ &
$\mathbb{1}_4 $ & $\mathbb{1}_4 $ & $ \eta^{e,0}_{\mu\nu}$
&$\mathbb{1}_4$&$\gamma_5\slashed{S}_{N}^f$& $\eta^{N,\vec s_N^{\,f}}_{\mu\nu}$\\
$ P_{N'}^{eN} $ &
$\gamma_5\slashed{S}_e^i $ &$\mathbb{1}_4$ & $\eta^{e,\vec s_e^{\,i}}_{\mu\nu}$&
 $\gamma_5\slashed{S}_N^i $&$\gamma_5\slashed{S}_{N}^f$& $\eta^{N,\vec s_N^{\,i}, \vec s_N^{\,f}}_{\mu\nu}$\\
\colrule
$ P_{e'N'}^e $ &
$\gamma_5\slashed{S}_e^i $ &$\gamma_5\slashed{S}_{e}^f$ & $\eta^{e,\vec s_e^{\,i},\vec s_e^{\,f}}_{\mu\nu}$&
 $\mathbb{1}_4 $&$\gamma_5\slashed{S}_{N}^f$& $\eta^{N,\vec s_N^{\,f}}_{\mu\nu}$\\
$ P_{e'N'}^N $ &
$\mathbb{1}_4$ &$\gamma_5\slashed{S}_{e}^f$ & $\eta^ {e,\vec s_e^{\,f}}_{\mu\nu}$
& $\gamma_5\slashed{S}_N^i$&$\gamma_5\slashed{S}_{N}^f$& $\eta^{N,\vec s_N^{\,i}, \vec s_N^{\,f}} _{\mu\nu}$\\
\end{tabular}
\end{ruledtabular}
\label{tab2}
\end{table}

\subsection{Differential scattering cross section}

The general differential cross section including the beam-target
asymmetries $A^{e,j;N,l}$ with respect to the initial electron and
nucleon spins $\vec s^{\, i}_e$ and $\vec s^{\, i}_N$, respectively,
has the form
\beq
\frac{d\sigma}{d\Omega_e}=\frac{d\sigma_0}{d\Omega_e}
\Big(1+\sum_{jl} s^i_{e; j} s^i_{N;l} A^{e,j;N,l}\Big)\,.
\eeq
The unpolarized cross section is given by
\beq
\frac{d\sigma_0}{d\Omega_e}=F_{kin}
\, \Sigma_{fi}^0\,,
\eeq
with
\beqa
\Sigma_{fi}^0&=&\sum_{\mu\nu}\eta_{\mu\nu}^{e,0}\,
\eta^{N,0;\mu\nu}\,,\nonumber\\
&=& \frac{1}{2^4m_e^2 m_N^2}\Big[ \frac{G_E^2+\tau G_M^2}{1+\tau}\,
\Big( (K\cdot P)^2 -Q^2 P^2\Big) +2G_M^2Q^4(1-\frac{2m_e^2}{Q^2})\Big] \,.
\eeqa
For the high energy limit and an intial nucleon at rest ($\vec p=0$) one obtains with
\beq
\widetilde F_{kin}=\frac{4\alpha^2 m_e^2 k^{\prime 2}}{Q^4 k^2}\,\quad
\mathrm{and}\quad
(K\cdot P)^2-Q^2P^2=4m_N^2Q^2\cot ^2\theta_e/2\,,
\eeq
and thus
\beq
\Sigma_{fi}^0=\frac{Q^2}{4m_e^2}\Big(\frac{G_E^2+\tau G_M^2}{1+\tau}\,
\cot ^2\theta_e/2+2\tau G_M^2\Big)\,,
\eeq
the high energy standard differential cross section for the lab frame
\beq
\frac{d\sigma_0}{d\Omega_e}=\frac{\alpha^2 k^{\prime 2}}{Q^2 k^2}
\Big(\frac{G_E^2+\tau G_M^2}{1+\tau}\,
\cot ^2\theta_e/2+2\tau G_M^2\Big)\,.
\eeq

The beam-target asymmetries are given by
\beqa
A^{e,j;N,l} &=&\frac{1}{\Sigma_{fi}^0}\,
\sum_{\mu\nu}\eta_{\mu\nu}^{e,\vec e_j}
\eta^{N,\vec e_l;\mu\nu}\,,
\eeqa
where $\vec e_j$ and $\vec e_l$ denote unit vectors of a chosen
coordinate system for a given reference frame.
Using the tensors of Eqs.\ (\ref{p-tensor-s}) and (\ref{e-tensor-s})
\beqa
\eta_{\mu\nu}^{e,\vec e_j}\eta^{N,\vec e_l;\mu\nu}
&=&
\frac{G_M}{2m_em_N}
\Big[G_E \Omega\Big (S_{e}^i(\vec e_j),S_{N}^i(\vec e_l) \Big)
-\frac{\tau  (G_M-G_E)}{(1+\tau)}\,
\Pi \Big (S_{e}^i(\vec e_j),S_{N}^i(\vec e_l) \Big)
\Big]\,,
\eeqa
where the notation 
\beqa
\Omega(S_{e},S_{N}) &=&
\Omega_{\mu\nu}S_{e}^\mu \,S_{N}^\nu\nonumber\\
&=&q^2\, S_e \cdot S_N- q\cdot S_e\, q\cdot S_N
\,,\label{QSS}\\
\Pi(S_{e},S_{N}) &=&
P\cdot S_{e} \,P\cdot S_{N}\,, \label{PSS}
\eeqa
has been introduced, one obtains
\beqa
A^{e,j;N,l}&=&
\frac{ G_M}{2m_em_N\Sigma^0_{fi}}\,
\Big[G_E \,
\Omega\Big(S_{e}^i(\vec e_j), S_{N}^i(\vec e_l) \Big)
- \frac{\tau(G_M-G_E)}{1+\tau}\, 
\Pi \Big( S_{e}^{i}(\vec e_j),  S_{N}^{i}(\vec e_l) \Big)
\Big]\,.\label{beam-target-asy}
\eeqa

In case that parity non-conservation is considered, two more vector
observables will appear, namely beam and target asymmetries $\vec A^{e}$
and $\vec A^{N}$, respectively,
\beq
\frac{d\sigma^{PV}}{d\Omega_e}=\frac{d\sigma_0}{d\Omega_e}
\Big(\vec s^{\,i}_e\cdot \vec A^e +\vec s^{\,i}_N\cdot \vec A^N\Big)\,.
\eeq

\subsection{Polarization of one of the final particles}

The polarization $\vec P_{a'}$ ($a'\in \{e',N'\}$) of one of the
outgoing particles $a'$ is governed by the polarization transfer from one
of  the initial particles $a\in \{e,N\}$ to the final one 
\beq
P_{a',j}\frac{d\sigma}{d\Omega_e}=\frac{d\sigma_0}{d\Omega_e}
\sum_l \Big(s^{i}_{e,l} P^{e,l}_{a',j} +
s^{i}_{N,l}P^{N,l}_{a',j} \Big)\,.
\eeq
Here 
$P^{e/N,l}_{a',j}$ denotes the polarization transfer from an
initial electron or nucleon polarized along $\vec e_l$ to the
polarization component along $\vec e_j$ of a final
particle $a'$. 

Polarization transfer $P^{e,l}_{N',j}$ from electron to nucleon has formally the same
structure as $A^{e,j;N,l}$ in Eq.~(\ref{beam-target-asy}) except for
the replacements $S_{e}^i(\vec e_j) \to S_{e}^i(\vec e_l)$ and 
$S_{N}^i(\vec e_l) \to S_{N}^f(\vec e_j)$. Thus it is given by
\beqa
P^{e,l}_{N',j}&=&\frac{1}{\Sigma_{fi}^0}\, \sum_{\mu\nu}
\eta_{\mu\nu}^{e, \vec e_{l}}
\eta^{N,\vec e_j;\mu\nu}\,,\nonumber\\
&=&
\frac{G_M }{2m_em_N\Sigma^0_{fi}}\,
\Big[G_E \, \Omega\Big(S_{e}^i(\vec e_l),S_{N}^{f}(\vec e_j)\Big)
- \frac{\tau(G_M-G_E)}{1+\tau}\, \Pi \Big(S_{e}^i(\vec e_l),S_{N}^f(\vec e_j) \Big)
\Big]\,.\label{e-N-pol-transf}
\eeqa
From this expression one obtaines the nucleon-electron spin transfer
by exchanging the initial and final states ($(e,i)\to 
(N,i)$ and  $(N,f)\to (e,f)$)
\beqa
P^{N,l}_{e',j}&=&\frac{1}{\Sigma_{fi}^0}\, \sum_{\mu\nu}
\eta_{\mu\nu}^{e, \vec e_{j}}
\eta^{N, \vec e_l;\mu\nu}\,,\nonumber\\
&=&
\frac{G_M}{2m_em_N\Sigma^0_{fi}}\,G_M
\Big[G_E \,\Omega\Big(S_{N}^i(\vec e_l),S_{e}^f(\vec e_j)\Big)
- \frac{\tau(G_M-G_E)}{1+\tau}\, \Pi\Big(S_{N}^i(\vec e_l),S_{e}^f(\vec e_j)\Big)
\Big]\,.\label{N-e-pol-transf}
\eeqa
As expected, the polarization transfer from electron to nucleon is
formally equivalent to the transfer from nucleon to electron.

The spin transfer from the initial to the final electron $P^{e,l}_{e';j}$ is given by
\beqa
P^{e,l}_{e';j}&=&\frac{1}{\Sigma_{fi}^0}\, \sum_{\mu\nu}
\eta_{\mu\nu}^{e, \vec e_{l},\vec e_j}
\eta^{N,0;\mu\nu}\,,\nonumber\\
&=&\frac{1}{2^4m_e^2m_N^2 \Sigma_{fi}^0}
\Big\{
\frac{G_E^2+\tau G_M^2}{1+\tau}\,
\Big[2P^2\,\Sigma_{e,2} (q) +2q^2\,\Sigma_{e,2} (P)
-\Big(q^2P^2+(K\cdot P)^2\Big) \Sigma_{e,0}\nonumber\\
&&
+2K\cdot P \,\,\Sigma_{e,2; \mu \nu}P^\mu K^\nu \Big]
-4m_e^2q^2 G_M^2  \Sigma_{e,0}\Big\}\,.
\label{e-e-pol-transf}
\eeqa

Similarly, the spin transfer from the initial to the final nucleon 
$P^{N,l}_{N';j}$ with $S_{N}^i=S_{N}^i(p, \vec e_{l})$ and 
$S_{N}^f=S_N^f(p',\vec e_{j})$  is 
\beqa
P^{N,l}_{N';j}&=&\frac{1}{\Sigma_{fi}^0}\, \sum_{\mu\nu}
\eta_{\mu\nu}^{e,0}
\eta^{N, \vec e_{l},\vec e_j;\mu\nu}\,,\nonumber\\
&=&
\frac{1}{2^4m_e^2m_N^2 \Sigma_{fi}^0}\Big[
-\Big\{4m_e^2q^2 G_M^2+\Big(P^2q^2+(K\cdot P)^2\Big)
\frac{ G_E^2+\tau G_M^2}{1+\tau}\Big\}
\Sigma_{N,0}
\nonumber\\
&&+2\Big\{4m_e^2 G_M^2+\frac{(K\cdot P)^2}{4m_N^2} 
\frac{(G_E- G_M)^2}{(1+\tau)^2}-q^2 \frac{G_E^2+\tau G_M^2}{1+\tau}
\Big\}
\Sigma_{N,2} (q)
\nonumber\\
&&
+2q^2\,\Sigma_{N,2}(K)\,G_M^2 + 2K\cdot P\,\, \Sigma_{N,2; \mu \nu}P^\mu K^\nu
\frac{G_M(G_E+\tau G_M)}{1+\tau}\Big]\,.
\label{N-N-pol-transf}
\eeqa

Again for parity
non-conservation one has $P^{0}_{a',j}$ and $P^{e,j;N,l}_{a',j}$ as
additional observables:
\beq
P_{a',j}^{PV}\frac{d\sigma}{d\Omega_e}=\frac{d\sigma_0}{d\Omega_e}
\Big (P^0_{a',j}+ \sum_{lr} s^i_{e,l} s^i_{N,r} P^{e,l;N,r}_{a',j}\Big)\,.
\eeq

\subsection{Spin correlations between both outgoing particles}

The spin correlations between the polarization components of both
outgoing particles are determined by two contributions
\beq
P_{e',l;N',j}\frac{d\sigma}{d\Omega_e}=\frac{d\sigma_0}{d\Omega_e}
\Big (P^0_{e',l;N',j}+ \sum_{rt} s^e_r s^N_t P_{e',l;N',j}^{e,r;N,t}\Big)\,.
\eeq
The first, $P^0_{e',l;N',j}$, denotes a spin correlations for an unpolarized
initial state and $P_{e',l;N',j}^{e,r;N,t}$, a beam-target asymmetry
of a spin correlation, if both initial
particles are polarized. $P^0_{e',l;N',j}$ is obtained from
$P^{e,l}_{N',j}$ in Eq.~(\ref{e-N-pol-transf}) replacing $S_e^i$ by
$S_e^f$, i.e.
\beqa
P^0_{e',l;N',j}&=&\frac{1}{\Sigma_{fi}^0}\, \sum_{\mu\nu}
\eta_{\mu\nu}^{e, \vec e_{l}}
\eta^{N,\vec e_j;\mu\nu}\,,\nonumber\\
&=&
\frac{G_M }{2m_em_N\Sigma^0_{fi}}\,
\Big[G_E \,\Omega\Big(S_{e}^{f}(\vec e_{l}), S_{N}^{f}(\vec e_{j})\Big)
- \frac{\tau(G_M-G_E)}{1+\tau}\, \Pi\Big(S_{e}^{f}(\vec e_{l}), S_{N}^{f}(\vec e_{j})\Big)
\Big]\,.\label{e-N-pol-corr}
\eeqa
This expression resembles formally the beam-target asymmetry of the
cross section in Eq.~(\ref{beam-target-asy}) replacing the initial by the final spins. 
The beam-target asymmetry of  the final spin correlation
$P_{e',l;N',j}^{e,r;N,t}$, is more complicated and thus listed in
Appendix~\ref{app2}. 

The parity violating part is determined by two contributions, 
$P^{e/N,r}_{e',l;N',j}$, a beam or target asymmetry of the final
spin correlation, if one of the initial particles is polarized, 
\beq
P_{e',l;N',j}^{PV}\frac{d\sigma}{d\Omega_e}=\frac{d\sigma_0}{d\Omega_e}
 \sum_r \Big (s^{i}_{e,r} P_{e',l;N',j}^{e,r} +
s^{i}_{N,r} P^{N,r}_{e',l;N',j} \Big)\,.
\eeq
This concludes the explicit presentation of the various parity
conserving polarization observables.
All above results are valid for any reference frame. 

As an application we consider in Appendix~\ref{app3} the case of
longitudinally polarized initial electrons for two interesting
observables, i.e.\ beam-target asymmetry and electron-nucleon
polarization transfer, with further specialization to lab and
Breit frames.  

\section{Conclusions\label{conclusion}}
In the present work we have derived explicit expressions for all
possible parity conserving observables of electron-nucleon elastic
scattering in the one-photon exchange approximation for the scattering matrix
without resorting to a special frame of reference. These expressions
are easily evaluated for a given frame of reference with a
corresponding choice of a coordinate system. The compact notation allows an easy
implementation into a computer code. Essentially, all 
results apply also to the inverse process of elastic nucleon scattering off
electrons. 

\begin{acknowledgments}
The author would like to thank the A1-collaboration and in particular
D.\ Izraeli,  J.\  Lichtenstadt, S.\  Paul,
E.\  Piasetzky, and  I.\  Yaron for the motivation and for many interesting discussions.
This work was supported by the Deutsche Forschungsgemeinschaft
(Collaborative Research Center 1044).
\end{acknowledgments}

\appendix
\section {Derivation of the leptonic and hadronic tensors \label{app1} }

The starting point for the evaluation of the tensors is
Eq.~(\ref{tensor}). First we consider the general form of a  Dirac
current for a nucleon 
\beqa
J^{fi}_{N,\mu}(p',s';p,s)&=&\bar u_N(p',s')\Big(A\gamma_\mu+\frac{B}{m_N}P_{\mu}
\Big)u_N(p,s)\,,
\eeqa
where $u_N(p,s)$ denotes a nucleon Dirac spinor, $P=p'+p$, and 
\beqa
A&=&G_M(Q^2) \,,\\
B&=& \frac{1}{2(1+\tau)}( G_E(Q^2)
-G_M(Q^2) )\,, 
\eeqa
with $\tau=Q^2/(4m_N^2)$, and electric ($G_E$) and magnetic
$(G_M)$ Sachs form factors as functions of the 
squared four momentum transfer $Q^2$. 

Introducing
\beqa
j_{N,\mu}(p',s';p,s)&=&\bar u_N(p',s')\gamma_\mu u_N(p,s)\,,
\\
\rho_N(p',s';p,s)&=&\bar u_N(p',s')u_N(p,s)\,,
\eeqa
one obtaines
\beqa
J^{fi}_{N,\mu}(p',s';p,s)&=&Aj_{N,\mu}(p',s';p,s)
+\frac{B}{m_N}P_{\mu} \rho_N(p',s';p,s)\,.
\eeqa
For the electron current one has to substitute $p\to k$, $p'\to k'$, 
$u_N(p/p',s/s')\to u_e(k/k',s/s')$, $A=1$, and $B=0$.

In view of the two terms of the current, one obtains three contributions
to the tensor $\eta^N_{\mu\nu}$
\beq
\eta^N_{\mu\nu}(p',p,{\cal O})=A^2 \bar\eta_{\mu\nu}(p',p;{\cal O}) 
+AB\,\widetilde\eta^N_{\mu\nu}(p',p;{\cal O}) 
+B^2\widehat\eta^N_{\mu\nu}(p',p;{\cal O})\,,
\eeq
where
\beqa
\bar\eta^N_{\mu\nu}(p',p;{\cal O})&=&\frac{1}{2}\,
\sum_{s' s \bar s' \bar s}
j_\mu(p',s';p,s)
\rho^i(p,\vec s^{\,i}; {\cal O})_{s \bar s} 
j_\mu ^\dagger (p',\bar s^{\,\prime};p,\bar s)
\rho^f (p',\vec s^{\,f}; {\cal O})_{\bar s^{\,\prime} s'}
\nonumber\,,\\
&=&\frac{1}{2}\,\sum_{s' s \bar s' \bar s}\bar u(p',s')\gamma_\mu u(p,s) 
\bar u(p,s){\cal S}_N^i ({\cal O})u(p,\bar s)
\bar u(p,\bar s)\gamma_\nu u(p',\bar s^{\,\prime})
\bar u(p',\bar s^{\,\prime}){\cal S}_N^f ({\cal O})u(p,s')\,.
\eeqa
The expressions for $\widetilde\eta^N_{\mu\nu}(p',p;{\cal O})$ and
$\widehat\eta^N_{\mu\nu}(p',p;{\cal O})$ will be presented below. 
Using the property
\beq
{\cal U}_N(p)=\sum_s u_N(p,s)\bar u_N(p,s)=\frac{1}{2m_N}\,(\slashed{p}+m_N)\,,
\eeq
the trace over the spin degrees of freedom becomes a trace over Dirac
matrices (indicated by the superscript D)
\beq
\bar\eta^N_{\mu\nu}(p',p;{\cal O})=
\frac{1}{2}\, Tr^D\Big(\gamma_\mu {\cal U}_N (p)
{\cal S}_N^i({\cal O}) {\cal U}_N (p)\gamma_\nu {\cal U}_N (p') 
{\cal S}_N^f({\cal O}){\cal U}_N (p')\Big)\,.
\eeq
The complete electron tensor $\eta^e_{\mu\nu}(k',k;{\cal O})$ is
obtained from this expression by the substitutions 
$(p,p')\to (k,k')$, ${\cal U}_N (p/p')\to {\cal U}_e (k/k')$, 
and ${\cal S}_N^{i/f}({\cal O})\to {\cal S}_e^{i/f}({\cal O})$, 
while for the nucleon tensor one has two further contributions, i.e.
\beqa
\widetilde\eta^N_{\mu\nu}(p',p;{\cal O})&=&\frac{1}{m_N}\,
\Big(P_\mu\tau_{\nu}(p',p;{\cal O})
+P_\nu\widetilde\tau_{\mu}(p',p;{\cal O})\Big)\,,\\
\widehat\eta^N_{\mu\nu}(p',p;{\cal O}) &=&
\frac{1}{m_N^2}\, P_\mu P_\nu\,\rho_0(p',p;{\cal O})\,,
\eeqa
with
\beqa
\tau^N_\mu(p',p;{\cal O})&=&\frac{1}{2}\, Tr^D\Big({\cal U}(p)
{\cal S}_N^i({\cal O}) {\cal U}(p)\gamma_\mu {\cal U}(p') 
{\cal S}_N^f({\cal O}){\cal U}(p')\Big)\,,\label{tau}\\
\widetilde \tau^N_\mu(p',p;{\cal O})&=&\frac{1}{2}\, Tr^D\Big({\gamma_\mu\cal U}(p)
{\cal S}_N^i({\cal O}) {\cal U}(p) {\cal U}(p') 
{\cal S}_N^f({\cal O}){\cal U}(p')\Big)\,,\nonumber\\
&=&\frac{1}{2}\, Tr^D\Big({\cal U}(p')
{\cal S}_N^f({\cal O}) {\cal U}(p')\gamma_\mu {\cal U}(p) 
{\cal S}_N^i({\cal O}){\cal U}(p)\Big)=\tau_\mu(p,p';{\cal
  O})|_{\vec s _N^{\,i}\leftrightarrow \vec s _N^{\,f}}\,, \label{tau-tilde}\\
\rho^N_0(p',p;{\cal O})&=&\frac{1}{2}\, Tr^D\Big({\cal U}(p)
{\cal S}_N^i({\cal O}) {\cal U}(p) {\cal U}(p') 
{\cal S}_N^f({\cal O}){\cal U}(p')\Big)\,. \label{rho}
\eeqa

According to Eq.~(\ref{rho_op_e}) (see also Table I) one has for
${\cal S}_{e/N}^{i/f}$ two possibilities 
depending on the observable, i.e.\ whether the corresponding initial
or final state is polarized or not, resulting in four types of tensors 
$\bar\eta^x_{\mu\nu}(p',p)$ 
($x \in \{0; \vec s^{\,i}; \vec s^{\,f}; \vec s^{\,i},\vec s^{\,f}\}$):
\beqa
\bar\eta^{N0}_{\mu\nu}(p',p)&=&\frac{1}{2}\, Tr^D\Big(\gamma_\mu {\cal U}(p)
{\cal U}(p)\gamma_\nu {\cal U}(p') {\cal U}(p')\Big)\,,\nonumber\\
&=&\frac{1}{4m _N^2}\,(P_\mu P_\nu -(p'-p)_\mu (p'-p)_\nu 
+ (p'-p)^2g_{\mu\nu})\,, \nonumber\\
&=&\frac{1}{4m _N^2}\,(P_\mu P_\nu +\Omega_{\mu\nu})
\label{bareta_0}\\
\bar\eta^{N,\vec s _N^{\,i}}_{\mu\nu}(p',p)&=&\frac{1}{2}\, Tr^D\Big(\gamma_\mu {\cal U}(p)
\gamma_5\slashed{S}_N^i
{\cal U}(p)\gamma_\nu {\cal U}(p') {\cal U}(p')\Big)\,,\nonumber\\
&=&\frac{i}{2m _N}\,\varepsilon_{\mu\nu\alpha\beta} (p'-p\,)^\alpha  S _N^{i;\beta}\,, 
\label{bareta_si}\\
\bar\eta^{N,\vec s _N^{\,f}}_{\mu\nu}(p',p)&=&
\frac{1}{2}\, Tr^D\Big(\gamma_\mu {\cal U}(p)
{\cal U}(p)\gamma_\nu {\cal U}(p') 
\gamma_5\slashed{S}_N^f {\cal U}(p')\Big)\,, \nonumber\\
&=&\frac{i}{2m _N}\,\varepsilon_{\mu\nu\alpha\beta}(p'-p\,)^\alpha S _N^{f;\beta}\,,
\label{bareta_sf}\\
\bar\eta^{N,\vec s_N^{\,i}, \vec s _N^{\,f}}_{\mu\nu}(p',p)&=&\frac{1}{2}\, Tr^D\Big(\gamma_\mu {\cal U}(p)
\gamma_5\slashed{S}_N^i
{\cal U}(p)\gamma_\nu {\cal U}(p') \gamma_5\slashed{S}_N^f 
{\cal  U}(p')\Big)\,,\nonumber\\
&=&\frac{1}{4m _N^2}\,\Big[
-\Big(2p'\cdot S _N^i\, p\cdot S _N^f+(p-p'\,)^2\,S _N^i \cdot S _N^f \Big)
g_{\mu\nu}
-2S _N^i \cdot S _N^f \,(p_\mu p^{\prime}_{\nu}+p_\nu p^{\prime}_{\mu})\nonumber\\
&&+(p-p'\,)^2\, (S^i_{N;\mu} S^f_{N;\nu} +S^i_{N;\nu} S^f_{N;\mu} )
+2p\cdot S^f (S^i_{N;\mu} p'_\nu+ S^i_{N;\nu} p'_\mu)
+2p'\cdot S^i (S^f_{N;\mu} p_\nu +S^f_{N;\nu} p_\mu)\Big]\,,\nonumber\\
&=&\frac{1}{4m_N^2}\,\Big[2\,\Sigma_{N,2}( q)\,g_{\mu\nu}
+\Sigma_{0}\,(\Omega_{\mu\nu}-P_\mu P_\nu) + q^2\Sigma_{N,2;\mu\nu}\nonumber\\
&&
+\Big ((P_\mu \Sigma_{N,2;\nu \rho}P^\rho
-q_\mu \Sigma_{N,2;\nu \rho}q^\rho) +(\mu \leftrightarrow \nu) \Big) \Big]
\,. \label{bareta_sisf}
\eeqa
One should note that $\bar\eta^{N,0}_{\mu\nu}(p',p)$ and
$\bar\eta^{N,\vec s^{\,i}, \vec s^{\,f}}_{\mu\nu}(p',p)$ are even under the 
interchange $\mu \leftrightarrow \nu$ while
$\bar\eta^{N,\vec s^{\,i}/\vec s^{\,f}}_{\mu\nu}(p',p)$ are odd. 
Furthermore, the interchange 
$i\leftrightarrow f $, i.e.\ $p \leftrightarrow p'$ and  $\vec s _N^{\,i}
\leftrightarrow \vec s _N^{\,f}$,
leaves $\bar\eta^{N,0}_{\mu\nu}(p',p)$ and 
$\bar\eta^ {N,\vec s^{\,i}, \vec s^{\,f}}_{\mu\nu}(p',p)$ unchanged and, finally, 
$\bar\eta^{N,\vec s^{\,i}}_{\mu\nu}(p',p)$ and $\bar\eta^{N,\vec s^{\,f}}_{\mu\nu}(p',p)$
have formally the same structure. These properties apply also to 
$\widetilde\eta^ {N}_{\mu\nu}(p',p;{\cal O})$
and $\widehat\eta^ {N}_{\mu\nu}(p',p;{\cal O})$.

We now will turn to the other two contributions
$\widetilde\eta^N_{\mu\nu}(p',p;{\cal O})$ and 
$\widehat\eta^N_{\mu\nu}(p',p;{\cal O})$, 
which in addition contribute to the hadronic
tensor only. Firstly  one obtains
for $\tau^N_\mu(p',p;{\cal O})$
\beqa
\tau^{N,0}_\mu(p',p)&=&\frac{1}{2m_N}\,P_\mu\,,\\
\tau^{N,\vec s^{\,i}/\vec s^{\,f}}_\mu(p',p)&=&\frac{i}{4m_N^2}\,
\varepsilon_{\mu\alpha\beta\gamma}\,S_N^{i/f;\alpha}\,q^\beta \,P^\gamma\,,\\
\tau^{N,\vec s^{\,i}, \vec s^{\,f}}_\mu(p',p)&=&
\frac{1}{2m_N}\,(\Sigma_{N,2;\mu\rho}P^\rho -\Sigma_{N,0}\,P_\mu)\,,
\eeqa
and 
\beqa
\rho^{N,0}_0(p',p)&=&\frac{1}{4m_N^2}\,P^2=1+\tau\,,\\
\rho^{N,\vec s^{\,i}}_0(p',p)&=&\rho^{N,\vec s^{\,f}}_0(p',p)=0\,,\\
\rho^{N,\vec s^{\,i}, \vec s^{\,f}}_0(p',p)&=&-\frac{1}{4m_N^2}\,(2\, \Sigma_{N,2} (q)
+\Sigma_{N,0}P^2)\,.
\eeqa
This then yields for $\widetilde\eta_{\mu\nu}(p',p;{\cal O})$ 
\beqa
\widetilde\eta_{\mu\nu}^{\,N,0}(p',p)&=&\frac{1}{m_N^2}\, P_\mu P_\nu\,,\\
\widetilde\eta_{\mu\nu}^{\,N,\vec s^{\,i}/\vec s^{\,f}}(p',p)
&=&\frac{i}{4m_N^3}\,(P_\mu
\varepsilon_{\nu \alpha \beta\gamma}
- P_\nu \varepsilon_{\mu \alpha \beta\gamma})
S^{i/f;\alpha}_N q^\beta P^\gamma\,,\\
\widetilde\eta_{\mu\nu}^{\,N,\vec s^{\,i}, \vec s^{\,f}}(p',p)&=&
\frac{1}{2m_N^2}\,\Big((P_\mu \Sigma_{N,2;\nu\rho}P^\rho +(\mu \leftrightarrow \nu))
-2\Sigma_{N,0}\,P_\mu P_\nu\Big)
\eeqa
Again $\widetilde\eta_{\mu\nu}^{\,N,0}$ and
$\widetilde\eta_{\mu\nu}^{\,N,\vec s^{\,i}, \vec s^{\,f}}$ are even and
$\widetilde\eta_{\mu\nu}^{\,N,\vec s^{\,i} /\vec s^{\,f}}$ are odd under the exchange
$\mu\leftrightarrow\nu$. For 
$\widehat\eta_{\mu\nu}(p',p;{\cal O})$ one finds
\beqa
\widehat\eta_{\mu\nu}^{\,N,0}(p',p)&=&\frac{P^2}{4m_N^4}\, P_\mu P_\nu
=\frac{1+\tau}{m_N^2}\,  P_\mu P_\nu\,,\\
\widehat\eta_{\mu\nu}^{\,N,\vec s^{\,i}}(p',p)
&=&\widehat\eta_{\mu\nu}^{\,N,\vec s^{\,f}}=0\,\\
\widehat\eta_{\mu\nu}^{\,N,\vec s^{\,i}, \vec s^{\,f}}(p',p)&=&
-\frac{1}{4m_N^4}\, \Big (2\,\Sigma_{N,2} (q)
+\Sigma_{N,0}\, P^2 \Big) P_\mu P_\nu \,.
\eeqa

\section{The beam-target asymmetries of the final spin 
  correlations\label{app2}}

Here we evaluate the more complex beam-target asymmetries of
the final spin  correlations:
\beqa
P_{e',l;N',j}^{e,m;N,n}&=&\frac{1}{\Sigma_{fi}^0}\, \sum_{\mu\nu}
\eta_{\mu\nu}^{e, \vec e_{m}, \vec e_{l}}
\eta^{N,\vec e_n,\vec e_j;\mu\nu}\,.
\eeqa
In view of the more involved operator structure of both tensors in
Eqs.~(\ref{p-tensor-ss}) and (\ref{e-tensor-ss}) we first write
them as sums over six operators $O_{\mu\nu}^{e/N,r}$ ($r=1,\dots,6$) with
corresponding coefficients $c_r^{e/N}$
\beqa
\eta^{N,\vec s_N^{\,i},\vec s_N^{\,f}}_{\mu\nu}
&=&\frac{1}{4m_N^2}
\,\sum_{r =1,6} c_r^N (\vec s_N^{\,i},\vec s_N^{\,f})\,
O_{\mu\nu}^{N,r}(\vec s_N^{\,i},\vec s_N^{\,f})\,,\\
\eta^{e,\vec s_e^{\,i},\vec s_e^{\,f}}_{\mu\nu}
&=&\frac{1}{4m_e^2}
\,\sum_{r =1,6} c_r^e (\vec s_e^{\,i},\vec s_e^{\,f})\,
O_{\mu\nu}^{e,r}(\vec s_e^{\,i},\vec s_e^{\,f})\,.
\eeqa
The operators $O_{\mu\nu}^{a,r}$ are symmetric four-tensors. They and
the corresponding coefficients are listed in Table~\ref{tab-op}. 
The notations
$\Omega_{\mu\nu}$, $\Sigma_{e/N,0}$, $\Sigma_{e/N,2}(v)$, and
$\Sigma_{e/N,2;\mu\nu}$ are defined in Eqs.~(\ref{qmunu}) through
(\ref{s2}). 
\begin{table}[h]
\caption{Listing of operators $O_{\mu\nu}^{a,r}$ and coefficients
  $c_r^a $ with $a\in \{e,N\})$.}
\begin{ruledtabular}
\begin{tabular}{l|l|r|l|r}
$r$ & $O_{\mu\nu}^{e,r}$ & $c_r^e $& 
$O_{\mu\nu}^{N,r}$ & $c_r^N $\\
\colrule
$1$  & $g_{\mu\nu}$ & $2\Sigma_{e,2}(q)$ & $g_{\mu\nu}$& $2G_M^2 \Sigma_{N,2}(q)$    \\
$2$ &  $\Omega_{\mu\nu}$ & $-\Sigma_{e,0}$ & $\Omega_{\mu\nu}$ & $-G_M^2 \Sigma_{N,0}$   \\
$3$ &  $K_\mu K_\nu$ & $-\Sigma_{e,0}$  & $P_\mu P_\nu$ &  
$-(G_E^2+\tau G_M^2) \Sigma_{N,0}/(1+\tau)$ \\
&&&&$-(G_M-G_E)^2 \Sigma_{N,2}(q) /(2m_N^2(1+\tau))$\\
$4$ &  $\Sigma_{e,2;\mu\nu}$ & $q^2$  &  $\Sigma_{N,2; \mu\nu}$  & $G_M^2 q^2$ \\
5 & $q_\mu \Sigma_{e,2;\nu\rho}q^\rho +(\mu\leftrightarrow\nu)$  & $-1$ &
$q_\mu \Sigma_{N,2;\nu\rho}q^\rho +(\mu\leftrightarrow\nu)$ & $-G_M^2$ \\
6 & $K_\mu \Sigma_{e,2;\nu\rho}K^\rho +(\mu\leftrightarrow\nu)$  & $1$ &
$P_\mu \Sigma_{N,2;\nu\rho}P^\rho +(\mu\leftrightarrow\nu)$ &
$G_M(G_E+\tau G_M)/(1+\tau)$\\
\end{tabular}
\end{ruledtabular}
\label{tab-op}
\end{table}

Then the contraction of the tensors reads
\beqa
\sum_{\mu\nu}
\eta_{\mu\nu}^{e, \vec s_e^{\,i}, \vec s_e^{\,f}}
\eta^{N,\vec s_N^{\,i},\vec s_N^{\,f};\mu\nu}
&=&
\sum_{r,s=1,6} c_r^e  c_s^N {\cal C}_{r,s}^{e,N}\,,
\eeqa
where the various operator contractions are denoted by
\beq
{\cal C}_{r,s}^{e,N} (\vec s_e^{\,i},\vec s_e^{\,f}; 
\vec s_N^{\,i},\vec s_N^{\,f})= 
O_{\mu\nu}^{e,r}(\vec s_e^{\,i},\vec s_e^{\,f})
O^{N,s;\mu\nu} (\vec s_N^{\,i},\vec s_N^{\,f})
\,.
\eeq
They are listed in Table~\ref{tab-con} with the additional notations
\beqa
\Sigma_{eN,0}&=&\Sigma_{e,2;\mu\nu}\Sigma_{N,2}^{\mu\nu}\,,\\
\Sigma_{eN,2}(v',v)&=&v'^\mu\Sigma_{e,2;\mu\nu}
\Sigma_{N,2}^{\nu\rho}v_\rho \,,\\
\Sigma_{eN,2}(v)&=&\Sigma_{eN,2}(v,v)\,.
\eeqa
\begin{table}[h]
\caption{Listing of contractions ${\cal C}_{r,s}^{e,N} $.}
\begin{ruledtabular}
\begin{tabular}{r|r|r|r|r|r|r}
$r\backslash s$ & 1 & 2 & 3 & 4 & 5 & 6 \\
\colrule
1 & 4 & $3q^2$ & $P^2$ & $2\Sigma_{N,0}$ & $2\Sigma_{N,2}(q)$ &  $2\Sigma_{N,2}(P)$
  \\
2 & $3q^2$ & $3q^4$ & $q^2P^2$ & $2(q^2\Sigma_{N,0}-\Sigma_{N,2}(q)$ & $0$ & $2q^2\Sigma_{N,2}(P)$ \\
3 & $K^2$ & $q^2 K^2$ & $(K\cdot P)^2$ & $\Sigma_{N,2}(K)$ & 0 &
$2K\cdot P\, \Sigma_{N,2}(K,P)$ \\
4 & $2\Sigma_{e,0}$ & $2(q^2\Sigma_{e,0}-\Sigma_{e,2}(q))$ & $\Sigma_{e,2}(P)$ &
$\Sigma_{eN,0}$ & $2\Sigma_{eN,2}(q)$ & $2\Sigma_{eN,2}(P)$ \\
5 & $2\Sigma_{e,2}(q)$ & 0 & 0 & $2\Sigma_{eN,2}(q)$  &
 $2(q^2\Sigma_{eN,2}(q)$ & $2\Sigma_{e,2}(q,P) \Sigma_{N,2}(q,P)$  \\
&&&&&$ + \Sigma_{e,2}(q) \Sigma_{N,2}(q))$&\\
6 & $2\Sigma_{e,2}(K)$ & $2q^2\Sigma_{e,2}(K)$ & 
$2K\cdot P\,\Sigma_{e,2}(K, P)$  & $2\Sigma_{eN,2}(K)$ &
 $2\Sigma_{e,2}(q,K) \Sigma_{N,2}(q,K)$ & $2(K\cdot P) \Sigma_{eN,2}(K,P)$\\
 &&&&&& $+\Sigma_{e,2}(K,P) \Sigma_{N,2}(K,P))$\\
\end{tabular}
\end{ruledtabular}
\label{tab-con}
\end{table}

Collecting the various
contributions, one obtains finally
\beq
P_{e',l;N',j}^{e,m;N,n}=\frac{1}{4^2 m_e^2 m_N^2\Sigma_{fi}^0}\, 
\sum_{r,s=1,6} c_r^e (\vec e_{m}, \vec e_{l})  c_s^N (\vec e_{n}, \vec e_{j})
{\cal C}_{r,s}^{e,N}(\vec e_{m}, \vec e_{l}; \vec e_{n}, \vec e_{j})\,.
\eeq

\section {Beam-target asymmetry and spin transfer for longitudinally polarized electrons at high energy\label{app3}}
In view of the recent analysis of final proton polarization
in electron scattering $\vec e \,(A,\vec p\,) A'$ in \cite{Ya17,Pa19},
we will now specialize to 
the case of electron to nucleon polarization transfer with
longitudinally polarized electrons at high energies. In addition, we
also consider the beam-target asymmetry of the cross section which is
formally similar.  

For longitudinally polarized electrons, i.e.\ 
$\vec s_e=\vec e_k=\vec k/\bar k$,
the spin vector $S_e$ has the form
\beq
S_e(k ,\vec e_k)=\frac{1}{m_e}(\bar k ,k_{0} \vec e_k)\,.\label{elongpol}
\eeq
According to Eq.~(\ref{espinb}) $ S_e$ has the Lorentz invariant property
\beq
 S_e(k ,\vec e_k)\cdot k=0\,.
\eeq
For electrons of sufficiently high
energy such that the electron mass can be neglected, i.e.\ 
$\bar k \gg m_e$ or $\bar
k\approx k_0$, the expression in 
eq.~(\ref {elongpol}) simplifies. In that case
one finds 
\beq
 S_e(k ,\vec e_k)\approx \frac{\bar k }{m_e} (1,\vec
e_k)\approx\frac{k}{ m_e}\label{e-spin-a} \,,
\eeq
which means the electron spin four vector
equals approximately its four momentum divided by its mass. 

Evaluating the expressions in eqs.~(\ref{QSS}) and (\ref{PSS})
\beqa
\Omega(\frac{k}{m_e},S_N^i)&=& \frac{1}{m_e}(q ^2 k\cdot S_N^i-k\cdot
q\,q\cdot S_N^i)\nonumber\\
&=&\frac{q^2}{2m_e}\,K\cdot S_N^i\,,\\
\Pi(\frac{k}{m_e},S_N^i)&=& -\frac{1}{m_e}\,k\cdot P\, q\cdot S_N^i\,,
\eeqa
where we have used
$k\cdot q=q^2/2$ and $P\cdot S_N^i=-q\cdot S_N^i$,
one obtains for the beam-target asymmetry 
(see Eq.~(\ref{beam-target-asy}))
\beqa
A^{e,long,N,j} &= &
\frac{1}{4m_e^2m_N\Sigma^0_{fi}}\,G_M
\Big[G_E \, q^2 \,K\cdot S_N^i (\vec e_j)
+2\frac{\tau(G_M-G_E)}{1+\tau}\, 
k\cdot P\, q\cdot S_N^i(\vec e_j)\Big]
\,,\label{beam-target-asy-long}
\eeqa
and for the polarization transfer component
(see Eq.~(\ref{e-N-pol-transf}))
$P^{e,long}_{N'; j} =P^{e,\vec   e_k}_{N'; j}$ 
\beqa
P^{e,long}_{N',j} &= &
\frac{1}{4m_e^2m_N\Sigma^0_{fi}}\,G_M
\Big[G_E \, q^2 \,K\cdot S_N^f (\vec e_j)
+2\frac{\tau(G_M-G_E)}{1+\tau}\, 
k\cdot P\, q\cdot S_N^f(\vec e_j)\Big]
\,.\label{e-N-lpol-transf}
\eeqa
As mentioned before these results are valid for any reference
frame. 

As special cases we will now consider these observables in the
lab and Breit frames, using the standard coordinate system~\cite{By78}, i.e.\ 
the $z$-axis along the three-momentum transfer 
$\vec q$, the $y$-axis along $\vec k\times\vec k^{\,\prime}$ 
and the $x$-axis to form a right-handed system.  The unit vectors will
be denoted by $\vec e_j$.

\subsection{The lab frame}
All lab frame quantities will be denoted by a subscript ``$L$''. 
Since the intial nucleon is at rest, the final nucleon momentum
is
\beq
p'_L=(E'_L,\vec q_L\,) \quad\mathrm{ with }\quad E'_L=\sqrt{m_N^2+\bar
  q^2_L}\,.
\eeq
Furthermore, from the Bjorken condition 
\beq
x_{Bj}=Q^2/2q\cdot p=Q^2/(E'_L-m_N)2m_N=1,
\eeq 
one finds $E'_L-m_N=Q^2/2m_N$, from which follows
\beq
\bar q_L^2=Q^2+(E'_L-m_N)^2=Q^2(1+\tau)\,.
\eeq
With the nucleon spin $\vec e_j$
in the nucleon's rest frame ($e_{qj}=\vec e_{q}\cdot \vec e_{j}$)
the initial and final nucleon spin four-vectors are 
\beqa
S^i_{N} (p_L,\vec e_{j})&=&(0, \vec e_{j} )\,\\
S^f_{N} (p'_L,\vec e_{j})&=&
\Big(\frac{\bar q_L}{m_N}e_{qj}, \vec e_{j} +2\tau e_{qj}\vec e_{q}\Big)
\,.
\eeqa
For the scalar products appearing in Eqs.~(\ref{beam-target-asy-long}) 
and (\ref{e-N-lpol-transf}) one has
\beqa
S^i_{N} (p_L,\vec e_j)\cdot q_L&=& -\bar q_L e_{qj}\,,\quad
S^i_{N} (p_L,\vec e_j)\cdot K_L=-\,\vec e_{j}\cdot\vec K_L\,,\\
 S^f_{N} (p'_L,\vec e_j)\cdot q_L&=& -\bar q_L e_{qj}\,, \quad
S^f_{N} (p'_L,\vec e_j)\cdot K_L=\frac{Q^2}{m_N\bar q_L}\,
(\bar k_L+\bar k_L')e_{qj}
-\,\vec e_{j}\cdot\vec K_L\,,
\eeqa
where we have used
\beq
\vec e_q\cdot \vec K_L=\frac{Q^2}{2m_N\bar q_L}\,(\bar k_L+\bar k'_L)\,.
\eeq
Then, using $\bar q_L=Q\sqrt{1+\tau}$ and 
$k_L\cdot P_L=m_N(\bar k_L+\bar k'_L)$,
one finds for the beam-target asymmetry and the electron-nucleon spin
transfer
\beqa
A^{e,long,N,j}_L&=&\frac{G_M}{4m_e^2m_N\Sigma_0}\,\Big[
G_EQ^2 \vec K_L\cdot\vec e_j
-(G_M-G_E)\frac{2m_N\tau Q}{\sqrt{1+\tau}}(\bar k_L+\bar k'_L)
e_{qj}\Big]
\,,\\
P^{e,long}_{N,j;L}&=&-\frac{G_M}{4m_e^2m_N\Sigma_0}\,\Big[
G_EQ^2 \Big (\frac{Q (\bar k_L+\bar k'_L)}{m_N \sqrt{1+\tau}}e_{qj}
-\vec K_L\cdot\vec e_j \Big)
+(G_M-G_E)\frac{2m_N\tau Q}{\sqrt{1+\tau}}(\bar k_L+\bar k'_L)
e_{qj}\Big]\nonumber\\
&=&A^{e,long,N,j}_L-\frac{G_MG_E}{m_e^2\Sigma_0}\,
\frac{\tau Q (\bar k_L+\bar k'_L)}{\sqrt{1+\tau}}e_{qj}
\,.
\eeqa
This gives for the $j=x$-component using $e_{qx}=0$ and
\beq
\vec e_{x}\cdot \vec K_L=2(\bar k_L \bar k'_L/\bar q_L)\sin\theta_L
=\frac{Q}{\sqrt{1+\tau}}\cot(\theta_L/2)
\eeq
with $\theta_L$ as scattering angle in the lab frame, 
\beqa
A^{e,long,N,x}_L=P^{e,long}_{N,x;L} &=&
\frac{m_N\tau Q \cot(\theta_L/2)}{m_e^2\Sigma_0 \sqrt{1+\tau}}\,
G_MG_E\,. \label{Ax-lab}
\eeqa
Correspondingly  for the $z$-component ($e_{qz}=1$ and $\vec
e_{z}\cdot \vec K_L=(\bar k_L+\bar k'_L)Q^2/(2m_N\bar q_L)$)
\beqa
A^{e,long,N,z}_L&=&
\frac{\tau Q (\bar k_L+\bar k'_L)}{2m_e^2\Sigma_0 \sqrt{1+\tau}}\,
G_M(2G_E-G_M)\,. \label{Az-lab}\\
P^{e,long}_{N,z;L} &=&-
\frac{\tau Q (\bar k_L+\bar k'_L)}{2m_e^2\Sigma_0 \sqrt{1+\tau}}\,
G_M^2\label{Pz-lab}
\,. 
\eeqa
Thus, one finds for the ratios of the $x$- over the $z$-components
\beqa
\frac{A^{e,long,N,x}_L}{A^{e,long,N,z}_L}
&=& \frac{2m_N}{(\bar k_L+\bar
  k'_L)\tan{(\theta_L/2)}}\,\frac{G_E/G_M}{2G_E/G_M-1}\,, \label{ratio-A-lab}\\
\frac{P^{e,long}_{N,x;L}}{P^{e,long}_{N,z;L}}&=&-
\frac{2m_N}{(\bar k_L+\bar
  k'_L)\tan{(\theta_L/2)}}\,\frac{G_E}{G_M}\,. \label{ratio-P-lab}
\eeqa
The latter result is well known. 
 Thus measuring the beam-target asymmetry is in a certain sense, i.e.\
 with respect to the ratio $G_E/G_M$, equivalent to a
measurement of the polarization transfer.

\subsection {The Breit frame}
The Breit or ``brick wall'' frame is defined by the condition $\vec
p^{\,\prime}_B=-\vec p _B\,$, which means $q_0=0$ and therefore
$Q^2=\bar q^2 _B$, and one finds, denoting all Breit frame quantities 
by a subscript ``$B$'',
\beq
p_B=(E_B,-\vec q _B/2),\quad p_B'=(E_B,\vec q
_B/2),\quad q_B=(0,\vec q_B\,),\quad P _B=(2E _B,\vec 0\,)\,, 
\eeq
where $E_B=m_N\sqrt{1+\tau}$. 
The initial and final nucleon spin four vectors with rest frame spin
in the direction of the unit vector $\vec s_{N}^{\,i/f}=\vec  e_{j}$ is given by 
\beq
S_{N}^{i/f} (p_B/p' _B,\vec e_{j})=
\Big(\mp \,\frac{\bar q _B e_{qj}}{2m_N}, 
\vec e_{j} +\frac{1}{m_N}\,(E_B-m_N)  e_{qj}\vec e_q\Big)\,.
\eeq
Then one obtains with $E_B=m_N\sqrt{1+\tau}$ and
$\bar q_B=Q$ and $\vec e_{q}\cdot \vec K_B=0$
\beqa
S_{N} ^{i/f} (p_B/p'_B,\vec e_j) \cdot q_B&=&-\frac{QE _B}{m_N} e_{qj}\,,\\
S_{N} ^{i/f}  (p_B/p'_B,\vec e_j) \cdot K_B&=&\mp \, \frac{Q\bar k_B}{m_N} \,
e_{qj} -\vec e_{j}\cdot \vec K_B\,.
\eeqa
With these expressions and $\vec k_B\cdot \vec P_B=2\bar k_BE_B$ one finds 
\beqa
A^{e,long,N,j}_B&=&\frac{m_N\tau}{m_e^2\Sigma_0}\,G_M\Big[
G_E\vec K_B\cdot\vec e_j
+(2G_E-G_M)\frac{Q \bar k_B}{m_N}
e_{qj}\Big]
\,,\\
P^{e,long}_{N,j;B}&=&-\frac{m_N\tau}{m_e^2\Sigma_0}\,G_M\Big[
G_E\vec K_B\cdot\vec e_j
-G_M\frac{Q \bar k_B}{m_N}e_{qj}\Big]\nonumber\\
&=&-A^{e,long,N,j}_B+\frac{2\tau Q \bar k_B e_{qj}}{m_e^2\Sigma_0}\,G_MG_E
\,.
\eeqa
This gives for the $j=x$-components using $e_{qx}=0$ and $\vec
e_{x}\cdot \vec K_B=2\bar k_B\cos{(\theta_B/2)}$ with $\theta_B$ as
scattering angle in the Breit frame 
\beqa
A^{e,long,N,x}_B=-P^{e,long}_{N,x;B}&=&\frac{2m_N\tau\bar k_B\cos{(\theta_B/2)}}
{m_e^2\Sigma_0}\,G_MG_E
\,,
\eeqa
and for the $j=z$-components with $e_{qz}=1$ and $\vec e_{z}\cdot \vec K_B=0$
\beqa
A^{e,long,N,z}_B&=&\frac{\tau Q\bar k_B}{m_e^2\Sigma_0}\,G_M
(2G_E-G_M)\,,\\
P^{e,long}_{N,z;B}&=&\frac{\tau Q\bar k_B}{m_e^2\Sigma_0}\,G_M^2
\,.
\eeqa
Thus the ratios of the $x$- over the $z$-components of the beam-target
asymmetry and the polarization transfer become
\beqa
\frac{A^{e,long,N,x}_B}{A^{e,long,N,z}_B}&=&
\frac{2m_N \cos(\theta_B/2)}{Q}\,\frac{G_E/ G_M}{2G_E/G_M-1}
\,, \label{ratio-A-breit}\\
\frac{P^{e,long}_{N,x;B}}{P^{e,long}_{N,z;B}}&=&-
\frac{2m_N \cos(\theta_B/2)}{Q}\,\frac{G_E}{G_M}\,. \label{ratio-P-breit}
\eeqa

The corresponding lab frame quantities can also be obtained from the
foregoing Breit frame ones by a Lorentz boost to the lab frame~\cite{Pu09}.

\end{document}